\documentclass[sigconf,natbib=true]{acmart}

\AtBeginDocument{%
  \providecommand\BibTeX{{%
    \normalfont B\kern-0.5em{\scshape i\kern-0.25em b}\kern-0.8em\TeX}}}

\copyrightyear{2024}
\acmYear{2024}
\setcopyright{acmlicensed}
\acmConference[SIGIR '24] {Proceedings of the 47th International ACM SIGIR Conference on Research and Development in Information Retrieval}{July 14--18, 2024}{Washington, DC, USA.}
\acmBooktitle{Proceedings of the 47th International ACM SIGIR Conference on Research and Development in Information Retrieval (SIGIR '24), July 14--18, 2024, Washington, DC, USA}
\acmISBN{979-8-4007-0431-4/24/07}
\acmDOI{10.1145/3626772.3657906}

\usepackage{nicefrac}
\usepackage{multirow}
\usepackage{xspace}
\usepackage{booktabs}
\usepackage{makecell}
\usepackage{moresize}
\usepackage{tabularx}
\usepackage{adjustbox}
\usepackage[normalem]{ulem}
\usepackage{siunitx}
\usepackage{tikz}
\usepackage{amsthm}
\usepackage{graphicx}
\usepackage{tablefootnote}

\usepackage{fontawesome5}

\newcommand{\github}[1]{%
\centering%
   \href{#1}{\faGithub} \url{#1}%
}

\newcommand{\rowgroup}[1]{\hspace{-1em}#1}

\newcommand\boldparagraph[1]{\vspace{0.35em}\noindent\textbf{#1}}

\newcommand{\deepimpact}{\textsf{DeepImpact}}

\newcommand{\msmarcodev}{\textsf{MSMARCO Dev Queries}\xspace}

\newcommand{\pulse}{\textsf{BMP}}

\newcommand{\splade}{\textsf{SPLADE}}
\newcommand{\esplade}{\textsf{ESPLADE}}
\newcommand{\unicoil}{\textsf{uniCOIL}}

\newcommand{\bmw}{\textsf{BMW}}
\newcommand{\maxscore}{\textsf{MaxScore}}
\newcommand{\clipping}{\textsf{Clipping}}
\newcommand{\ioqp}{\textsf{IOQP}}
\newcommand{\anytime}{\textsf{Anytime}}

\begin{document}

\title[Faster Learned Sparse Retrieval with Block-Max Pruning]{Faster Learned Sparse Retrieval with Block-Max Pruning}

\author{Antonio Mallia}
\affiliation{%
  \institution{Pinecone, Italy}
  \country{}}

\author{Torsten Suel}
\affiliation{%
 \institution{New York University, USA}
 \country{}}

\author{Nicola Tonellotto}
\affiliation{%
  \institution{University of Pisa, Italy}
  \country{}}


\renewcommand{\shortauthors}{Antonio Mallia \and Torsten Suel \and Nicola Tonellotto}

\begin{abstract}

Learned sparse retrieval systems aim to combine the effectiveness of contextualized language models with the scalability of conventional data structures such as inverted indexes. 
Nevertheless, the indexes generated by these systems exhibit significant deviations from the ones that use traditional retrieval models, leading to a discrepancy in the performance of existing query optimizations that were specifically developed for traditional structures.
These disparities arise from structural variations in query and document statistics, including sub-word tokenization, leading to longer queries, smaller vocabularies, and different score distributions within posting lists.

This paper introduces Block-Max Pruning (\pulse), an innovative dynamic pruning strategy tailored for indexes arising in learned sparse retrieval environments. \pulse\ employs a block filtering mechanism to divide the document space into small, consecutive document ranges, which are then aggregated and sorted on the fly, and fully processed only as necessary, guided by a defined safe early termination criterion or based on approximate retrieval requirements.
Through rigorous experimentation, we show that \pulse{} substantially outperforms existing dynamic pruning strategies, offering unparalleled efficiency in safe retrieval contexts and improved trade-offs between precision and efficiency in approximate retrieval tasks.

\vspace{5px}
\github{https://github.com/pisa-engine/BMP}            
\end{abstract}



\maketitle

\section{Introduction}

\begin{figure}[t]
\includegraphics[width=\linewidth]{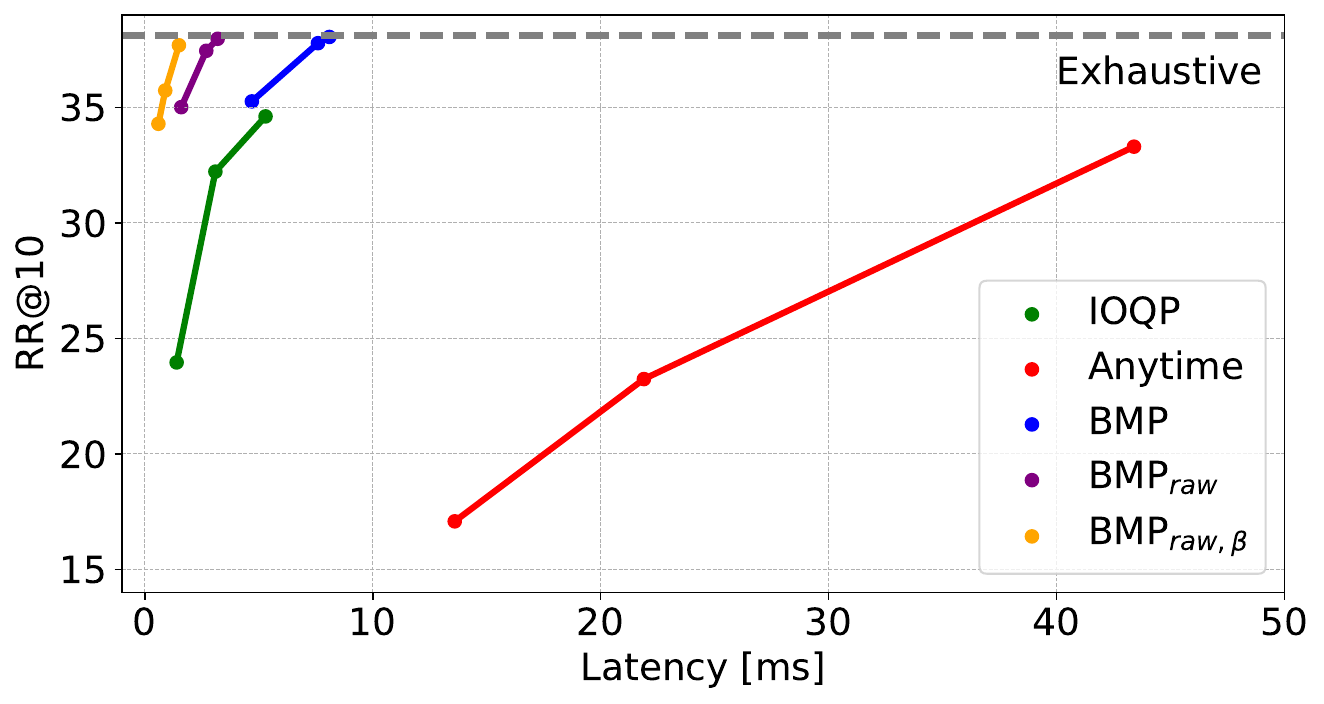}
\caption{Effectiveness-efficiency graph for different query processing algorithms. Every line corresponds to an algorithm, every point corresponds to a different configuration.}
\vspace{-5mm}
\label{fig:block-time}
\end{figure}

Information retrieval (IR) systems have increasingly turned towards learned sparse models for their ability to efficiently handle the complexity of query and document representations used in state-of-the-art ranking methods~\cite{unicoil,splade,deepimpactv1}. Learned sparse models operate by mapping queries and documents into sparse vectors, and then leveraging the traditional inverted index -- a data structure that is still crucial for scalable text retrieval~\cite{fntir2018}. This use of inverted indexes gives learned sparse model an advantage over dense vector search methods~\cite{star,colbert}, leading to faster query processing and reduced storage requirements.

Within the domain of learned sparse models, \splade{}~\cite{splade} and \unicoil{}~\cite{unicoil} have emerged as significant advancements. \splade{} operates on the principle of sparse lexical decomposition, effectively capturing the nuanced relationship between terms and documents. In contrast, \unicoil{}
incorporates token-level contextualization into the sparse modeling process. These models differ from methods such as \deepimpact~\cite{deepimpactv1}, which focuses more on direct impact score predictions for terms in (suitably expanded) documents.
Among the learned sparse models that have been proposed, we focus on \splade{}~\cite{splade} and \unicoil{}~\cite{unicoil} for their good effectiveness, and aim at optimizing their retrieval efficiency.
%
Learned sparse models are used to produce inverted indexes, a data structure which contains one inverted list for each distinct term, where each inverted list describes which documents are relevant to the term and how relevant they are. In particular, each inverted list consists of postings of the form $(d_i, s_i)$ where $d_i$ is the document ID (docID) of a document relevant to the term, and $s_i$ is an {\em impact score}, quantized to a limited number of bits, that models the degree of relevance of $d_i$.
Given an inverted index, the score of a document $d$ with respect to a query $q$ is defined as $s(q, d) = \sum_{t \in q} w(t, q) \cdot s(t, d)$ where $w(t, q)$ is a term weight for $t$ in $q$, and $s(t, d)$ is the corresponding impact score stored in the inverted list for $t$ -- or $0$ if there is no such posting for document $d$. Our goal is to return the top-$k$ scoring documents for a query, say for $k=10$ or $k=1000$.

Once the learned sparse model has been encoded into an inverted index, algorithms such as \maxscore~\cite{maxscore} and BlockMaxWand (\bmw)~\cite{bmw,vbmw,chaudhuri_ICDE01}, or score-at-a-time (SaaT) traversal algorithms~\cite{anh2001vector}, can be used to perform efficient query processing, by performing dynamic pruning during index traversal. However, the efficiency of these algorithms on learned sparse models does not match that commonly reported on traditional similarity models~\cite{wand-magic}, such as BM25 or query likelihood~\cite{bm25}. This has motivated several recent approaches that get better performance for index structures based on learned sparse models. This includes the \anytime{} approach from \citet{anytime}, which allows existing algorithms to operate on confined segments of a topically-clustered inverted index, and \clipping~\cite{clipping}, where \maxscore\ is adapted to enable efficient
query processing for learned term importance
schemes via a simple technique called term impact decomposition. Another set of approaches \cite{gt,2GTI2023} uses standard models such as BM25 for index traversal and then immediately reranks results with the learned sparse model.


This paper introduces \pulse, a novel query processing strategy that is optimized for the indexes generated by learned sparse models. \pulse\ employs a block filtering mechanism~\cite{sigir2013,live-block} to prioritise clusters of documents based on their potential relevance.
Promising subsets of documents that need to be processed are identified through an optimized computation of range-based upper bounds that are compared to a current threshold. When scoring a range of documents, \pulse\ relies on a variant of a forward index, 
i.e., an array of documents stored as term-impact pairs.


Our experiments on three learned sparse models, \splade{}, \esplade{}, and \unicoil{}, show that \pulse\ is significantly faster than previous methods for safe retrieval of top results, ranging from ${\sim}2\times$ to ${\sim}60\times$, while achieving a much better trade-off between efficiency and effectiveness when used for approximate query processing as shown in Figure~\ref{fig:block-time} for the MS MARCO dataset.

\vspace{-3mm}

\section{Block-Max Pruning}


We now describe our approach, called Block-Max Pruning (\pulse{}), which consists of two major phases, block filtering, and  evaluation of individual candidate blocks until some termination condition is satisfied. 

\boldparagraph{Block Filtering.} In this phase, we adapt the block filtering approach in \cite{sigir2013,live-block} to our scenario. Recall that each document is identified by a docID between $0$ and $n-1$ where $n$ is the collection size. We divide this range into blocks of $b$ consecutive docIDs, for some $b$ to be chosen later. Now for each term in the collection, we have $\lceil n/b \rceil$ blocks, and for each such block we store the maximum impact score of any posting in the block. (This may be $0$ if there are no postings for the term in the block.) Thus, for each term, we now have an array of $\lceil n/b \rceil$ block-max impact scores.

Given a query $q$, we can compute upper bounds for the possible scores of all documents by simply taking the arrays of block-max impact score for the query terms $t \in q$, and adding them up, weighted by term weights $w(t, q)$. As shown in \cite{sigir2013,live-block}, this can be done in a highly efficient manner using vectorized instructions available in modern CPUs. This gives us a  score upper bound, with respect to $q$, for each block of $b$ documents in the collection. 

\boldparagraph{Evaluating Candidate Blocks.} Next, we evaluate individual blocks of $b$ documents, in decreasing order of the block upper bounds, until some termination condition is met. To access blocks in decreasing order, we need to first sort the blocks by their upper bounds, or at least partially sort them. 

To efficiently evaluate a block, we use a hybrid between inverted and forward index. That is, for each block of $b$ documents, we store inverted lists of postings of the form $(d, s)$ where $d$ is a $\log_2 (b)$-bit number identifying a document within its block, and $s$ is the associated impact score, plus a sorted list of all terms that occur in the block, each with a pointer to the start of the corresponding posting list. We evaluate a block by fetching the postings for the query terms, and aggregating the impact scores into an array of $b$ accumulators. We use a heap to retain the $k$ highest-scoring documents found.

We stop evaluating blocks once a termination condition is met.

\boldparagraph{Representing Block-Max Impact Scores.}
Our approach uses fairly small blocks, in some cases as small as $b=8$ or $b=16$. Thus we have to store and aggregate fairly long arrays of $8$-bit impact scores -- for MS MARCO with $8.8M$ documents and $b=16$ we get more than $500K$ blocks and scores per distinct term. However, many scores are zero, and thus we use a sparse representation of the non-zero values.

\boldparagraph{Document Ordering.} The assignment of docIDs to documents can have a major impact on index size and processing speed in many scenarios, including ours here. We assign docIDs according to BP ordering \cite{dhulipala2016compressing}, considered state of the art in the area. This leads to sparser and thus more compressible block-max impact scores, and to tighter upper bounds for blocks after aggregation, as documents with similar scores are grouped into one block.

\boldparagraph{Partial Sorting.} After computing block upper bounds, we need to support block access in sorted order. Doing a full sort of the hundreds of thousands of upper bound scores would be very expensive. Instead, we use a simple top-$k$ threshold estimator based on single-term quantiles \cite{mallia2020comparison}, and sort only the block scores above the threshold using a simple counting sort-based approach.

\boldparagraph{Forward or Inverted Index.} We experimented with several index organizations for evaluating a block, including a standard inverted index, a forward index, and the hybrid structure discussed above which performed best. The inverted index resulted in many expensive pointer movements where we jump from block to block (both forward and backward), while a forward index requires us to separately locate the relevant postings in each document. Thus, our approach does not require any standard inverted index at all.

\boldparagraph{Termination Conditions.}  If we process blocks until the top-$k$ threshold in the heap is higher than the upper bound of the next block, we are guaranteed to retrieve the top-$k$ highest scoring results. We can also choose more aggressive approximate policies where we only process a certain number of blocks, or terminate if the current threshold is within a certain factor of the upper bound of the next block. 
In our approach, the granularity of approximation can be precisely controlled through an adjustable parameter, $\alpha$ (ranging between $0$ and $1$), which determines the early stopping condition. Specifically, processing of blocks continues until the top-$k$ threshold in the heap exceeds the $\alpha$-adjusted upper bound of the next block. This mechanism ensures that we can retrieve the top-$k$ highest scoring results with a tunable level of approximation.

In contrast to other approximate methods like SaaT where documents could be partially scored, our approach maintains the integrity of exact document scoring, enabling more accurate and reliable ranking of documents.

\boldparagraph{Query Term Pruning.}
Another approximation approach we explored is represented by query term pruning, which involves the removal of certain query terms when their weight is below a given threshold. For this, we make our system tunable using a $\beta$ parameter to set the percentage of query terms we want to drop.

\section{Experimental Results}
We assess the performance of the suggested approach by using the well-known MS MARCO Passage~\cite{msmarco} dataset.
The effectiveness and efficiency of query processing are evaluated by comparing all models against the \msmarcodev. The experiments were carried out in memory, utilizing a single thread on a Linux system equipped with dual 2.8 GHz Intel Xeon CPUs and 512 GiB of RAM.

\boldparagraph{\splade}\footnote{https://huggingface.co/naver/splade-cocondenser-ensembledistil} refers to the model named CoCondenser-EnsembleDistil in \cite{splade-distill}, the most effective \splade\ model, where the model has been initialized from a pretrained CoCondenser \cite{cocondenser} checkpoint and fine-tuned using knowledge distillation and hard negatives.

\boldparagraph{\esplade{}}\footnote{https://huggingface.co/naver/efficient-splade-V-large-doc} denoted as ESPLADE-V-large in \cite{esplade}, represents a more efficient iteration of \splade{}. Although effective on the MS MARCO dataset, it demonstrates limited capability in generalization compared to its predecessor, as on the BEIR dataset~\cite{splade-distill,esplade}.

\boldparagraph{\unicoil{}}~\cite{unicoil,unicoil-tilde}  assigns scalar weights to terms, instead of the vector weights used in the original COIL \cite{coil} formulation, and is further enhanced by incorporating a TILDE~\cite{tilde} expansion component.

\begin{table}[h]
\caption{Index space consumption (in GB) for \splade{}.}\vspace{-3mm}
    \centering
\begin{tabular}{lcccccc}
\toprule
\multicolumn{1}{c}{Block size} & $8$ & $16$& $32$& $64$ & $128$& $256$ \\

\midrule
Forward Index & 7.1 & 6.0 & 5.1 & 4.3 & 3.7 & 3.3 \\
\rowgroup{BM Index} &  \\
Raw & 30.0 & 15.0 & 7.4 & 3.7 & 1.9 & 0.9 \\
Compressed & 5.5 & 4.1 & 3.0 & 2.3 & 1.7 & \phantom{-} 1.2 \tablefootnote{Compression introduces some overhead, which is well amortized for small blocks but becomes significant for larger ones.} \\


\bottomrule
\end{tabular}
\label{tab:index-size}
\end{table}

We use Anserini \cite{anserini} for creating the inverted indexes of the collections, export them into the common index file format (CIFF) \cite{lin2020supporting}, and reorder using BP \cite{mackenzie2021faster, mackenzie2019compressing}. We run \maxscore{}~\cite{maxscore} and \bmw~\cite{bmw} using PISA \cite{pisa}. \bmw{} uses a block size of 40 postings.
For \anytime{} \footnote{https://github.com/JMMackenzie/anytime-daat} \cite{anytime}, \clipping{}\footnote{https://github.com/JMMackenzie/postings-clipping} \cite{clipping}, and \ioqp{}\footnote{https://github.com/JMMackenzie/IOQP} (an efficient implementation of SaaT) \cite{ioqp} we leverage the official reference implementations. 
\anytime{} uses \maxscore{} as its inner DaaT traversal algorithm and the index is split into $100$ clusters following similar experimentation in \cite{wacky2}. We also tested \anytime{} with \bmw{}, but do not report the results as they are worse than with \anytime\ coupled with \maxscore.
We experimented with the guided traversal method presented in \cite{2GTI2023} which can only perform approximate retrieval, but we decided not to include the results since the fastest version~\textsf{2GTI-Fast} resulted in longer running times than the slowest of our baseline methods in Table~3 (45.0 ms for \splade).

\pulse\ is written in Rust, and was compiled with rustc 1.77 using -O3 optimization as per the
default release profile. \pulse$_{raw}$ indicates that the BM-index is stored uncompressed, $b$ refers to the block size used to generate the BM-index, and $\alpha$ and $\beta$ control early stopping and query term pruning, respectively.






\begin{table*}
\caption{Query times (in ms) of different exact query processing strategies for $k={10, 100, 1000}$}\vspace{-3mm}
\centering
\begin{tabular}{lrrrrrrrrrrrrrrrrrr}
\toprule
Strategy & \multicolumn{6}{c}{\splade} &  \multicolumn{6}{c}{\esplade} &  \multicolumn{6}{c}{\unicoil} \\ 
 & \multicolumn{2}{c}{$k=10$}& \multicolumn{2}{c}{$k=100$} & \multicolumn{2}{c}{$k=1000$} & \multicolumn{2}{c}{$k=10$} & \multicolumn{2}{c}{$k=100$}& \multicolumn{2}{c}{$k=1000$} & \multicolumn{2}{c}{$k=10$} & \multicolumn{2}{c}{$k=100$}& \multicolumn{2}{c}{$k=1000$} \\ 
\midrule
\maxscore & 120.6 &  \scriptsize{(11.5x)}       & 152.8 & \scriptsize{(9.6x)} & 193.8  & \scriptsize{(7.0x)} & 13.6 & \scriptsize{(5.9x)} &  19.2 & \scriptsize{(4.2x)} & 28.2 & \scriptsize{(2.6x)} & 14.5 & \scriptsize{(5.8x)} & 19.5 & \scriptsize{(4.1x)} & 28.8 & \scriptsize{(2.9)}\\
\bmw  & 614.2 & \scriptsize{(58.5x)}  & 658.7 & \scriptsize{(41.4x)}  & 686.7 & \scriptsize{(24.9x)} & 10.8 & \scriptsize{(4.7x)} & 16.0 & \scriptsize{(3.5x)}& 27.0 & \scriptsize{(2.5x)} & 12.4 & \scriptsize{(5.0x)} & 18.4 & \scriptsize{(3.8)} & 31.7 & \scriptsize{(3.2)} \\
\ioqp  &     79.1   &  \scriptsize{(7.5x)} & 80.2 & \scriptsize{(5.0x)}& 80.8 & \scriptsize{(2.9x)} &     27.3  & \scriptsize{(11.9x)} & 26.9  & \scriptsize{(5.9x)} & 27.6 & \scriptsize{(2.5x)}& 34.8 & \scriptsize{(13.9x)} & 33.5 & \scriptsize{(7.0x)} & 35.5       & \scriptsize{(3.6x)} \\

\anytime  &   80.6 & \scriptsize{(7.7x)}     & 114.0 & \scriptsize{(7.2x)}&    163.1 & \scriptsize{(5.9x)} &   8.2    &   \scriptsize{(3.6x)}      &  12.6 & \scriptsize{(2.7x)} & 20.8 & \scriptsize{(1.9x)} & 8.3 & \scriptsize{(3.3x)} & 12.4 & \scriptsize{(2.6x)} & 19.4 & \scriptsize{(2.0x)}\\
\clipping  &  245.9 & \scriptsize{(23.4)} & 358.8 & \scriptsize{(22.6x)} & 504.1  & \scriptsize{(18.3x)} & 9.8  & \scriptsize{(4.3x)} & 15.0  & \scriptsize{(3.3x)}& 24.8  & \scriptsize{(2.3x)}&  9.1 & \scriptsize{(3.6x)}&  14.5 & \scriptsize{(3.0x)} &   25.6 & \scriptsize{(2.6x)} \\
\midrule
\rowgroup{\pulse} \\
$b=32$  & \textbf{10.5} && 23.1 & \scriptsize(1.5x) & 66.9 & \scriptsize(2.4x)& \textbf{2.3} && 6.6 & \scriptsize{(1.4x)}& 26.4 & \scriptsize{(2.4x)} & \textbf{2.5} & & 6.1 & \scriptsize{(1.3x)} & 21.1 & \scriptsize{(2.1x)}\\
$b=16$  & 11.0 & \scriptsize{(1.1x)}        & \textbf{15.9} &  & 37.8 & \scriptsize{(1.4x)}       & 2.6 &\scriptsize{(1.1x)}  &  \textbf{4.6} & & 15.1 & \scriptsize{(1.4x)} & 3.2 & \scriptsize(1.3x) & \textbf{4.8} & &12.9      &  \scriptsize{(1.3x)} \\
$b=8$  &  15.0 & \scriptsize{(x1.4)}        &  16.9 & \scriptsize(1.1x) & \textbf{27.6} &    &4.2& \scriptsize(1.8x)& 5.1 & \scriptsize{(1.1x)} & \textbf{10.9}   & & 5.0 & \scriptsize(2.0x)& 5.7 & \scriptsize{(1.2x)}&  \textbf{9.9} & \\
\bottomrule
\end{tabular}
\label{tab:overall}
\end{table*}

\boldparagraph{Index Size.} Table~\ref{tab:index-size} provides a summary of the space required (in GB) to store both the block-based forward index and the additional block-max index structure necessary for implementing cluster-based pruning with \splade, across various block sizes. As the block size increases, the size of the forward index consistently decreases. This reduction can be attributed to the fact that larger blocks facilitate more term overlap across documents within the same block, thus reducing redundancy.
Regarding the block-max index, it is observed that smaller blocks lead to increased memory requirements. However, when compressed, the extra space needed is less than the forward index size.

We do not compare index sizes between our method and baseline approaches, as the size differences between the inverted and forward indexes are generally minor in the broadest cases (for instance, the uncompressed inverted index for \splade\ is approximately 7.9 GB). While both types of indexes can undergo similar compression techniques, such discussions are beyond the scope of this paper. Instead, our primary focus is on the additional space needed by our filtering structure. It is important to highlight that for many techniques, such as \maxscore, the extra space requirement is minimal. In contrast, \bmw\ necessitates storing an upper bound score for each posting block, leading to higher space consumption.

\boldparagraph{Overall Comparison.} Next, we compare all methods in terms of overall efficiency, for different retrieval depths, and summarize the results in Table~\ref{tab:overall}. We observe that \splade{} is dramatically slower than \esplade{} and \unicoil{} for all the strategies, in particular for \bmw, \maxscore\ and \clipping. This is due to the fact that \splade{} strongly leverages query expansion, i.e. it uses queries with a large number of terms, which these algorithms are particularly sensitive to. Conversely, \anytime\ and \ioqp\ exhibit greater robustness to the variability in query length. \anytime\ benefits from operating on specific clusters of the index with shorter posting lists, whereas \ioqp\ adopts a more brute-force, term-at-a-time approach during safe searches, simplifying the retrieval process by eliminating the need for additional management of long queries.
Among the methods evaluated, \pulse\ stands out as the fastest on \splade{}, achieving speeds 2.9 to 7.5 times faster than its closest competitor across the three retrieval depths examined.

\esplade{} and \unicoil{}, while not employing explicit query expansion, present their own challenges: (i) potentially longer queries due to Transformer-based tokenizers, (ii) extended posting lists from additional document expansion, and (iii) skewed score distributions from model fine-tuning.
In this context, \bmw\ manages to outperform \maxscore, yet it faces challenges when compared to the competitive performances of \anytime\ and \clipping. \ioqp\ maintains nearly constant latency across different retrieval depths. \pulse\ demonstrates similar efficiency for both \esplade\ and \unicoil{}, consistently outperforming other strategies by at least a factor of two, particularly at the lowest $k$ values.


\begin{table}[h]
\caption{Query times (in ms) and RR@10 of different approximate query processing strategies for $k={10}$.}\vspace{-3mm}
\centering
\setlength{\tabcolsep}{3pt}
\begin{tabular}{l@{}rrrrrr}
\toprule
Strategy & \multicolumn{2}{c}{\splade} &  \multicolumn{2}{c}{\esplade} &  \multicolumn{2}{c}{\unicoil} \\ 
& \multicolumn{1}{c}{MRT} & \multicolumn{1}{c}{RR@10}  & \multicolumn{1}{c}{MRT} & \multicolumn{1}{c}{RR@10}  &\multicolumn{1}{c}{MRT} & \multicolumn{1}{c}{RR@10}  \\ 
\midrule
Exhaustive   & -- & 38.11 & -- & 38.81 & -- &  35.03\\
\midrule
\ioqp$_{1\%}$  & 1.4 & 23.96 & 1.2 & 27.62  & 1.3 & 24.30  \\
\ioqp$_{5\%}$  & 3.1 & 32.22 & 3.0 & 34.62  & 2.8 & 31.66  \\
\ioqp$_{10\%}$  & 5.3 & 34.61 & 5.4 & 36.63  & 4.9 & 33.60  \\
\anytime$_{5}$  &     13.6&  17.08 &  1.4 & 21.48 & 1.3 & 21.31\\
\anytime$_{10}$  &     21.9   & 23.24  &2.2 & 28.96& 2.1 & 28.22  \\
\anytime$_{30}$  &  43.4      & 33.30  & 4.4 & 36.74 & 4.3 & 34.08  \\
\midrule
\rowgroup{\pulse} \\
${b=256, \alpha=0.60}$   & 4.7 & 35.26 & 0.6 & 30.90 & 0.6 & 27.95\\
${b=128, \alpha=0.75}$   & 7.6 & 37.78 & 1.1 & 36.66& 1.1 & 33.14\\
${b=64, \alpha=0.85}$   & 8.1 & 38.05 & 1.5 & 38.22 & 1.5 & 34.70\\
\midrule
\rowgroup{+ raw BM Index} \\
${ b=64, \alpha=0.65}$   & 1.6 & 35.01 & 0.4 & 33.96 & 0.4 & 30.29 \\
${ b=64, \alpha=0.75}$   & 2.7 & 37.45 & 0.6 & 36.52 & 0.5 & 33.13  \\
${b=32, \alpha=0.85}$   & 3.2 & 37.97 & 0.7 & 38.11 & 0.7 & 34.63 \\
\bottomrule
\end{tabular}
\label{tab:approximate}
\vspace{-5mm}
\end{table}

\boldparagraph{Approximate Retrieval.}
Our next experiment involves approximate retrieval. \ioqp{} can be turned into an approximate method by setting the maximum number of postings to process per query; \citet{ioqp} propose to do that by finding that number as a fraction of the total number of documents in the collection. For this experiment we use ${1\%,5\%, 10\%}$ as fractions.
\anytime{} can be tuned by setting the maximum number of clusters that are accessed per query or by setting a query latency budget. For the maximum number of cluster, we use $5$, $10$ and $30$ out of the $100$ total clusters in the index. Although we experimented with a latency budget by aligning it with one of the execution times from \pulse{} tests, we found these results less compelling for the comparison.

Table~\ref{tab:approximate} shows the latency (in ms) for retrieving the top-$10$ documents in an unsafe scenario. \ioqp\ is the fastest method for \splade, but it consistently reduces the effectiveness across the different models w.r.t. the exhaustive, i.e., safe, scenario. Similarly, \anytime\ negatively impact the effectiveness, while being similarly efficient as \ioqp\ for \esplade\ and \unicoil, resulting in the worst trade-off.
On the other side, \pulse\ is able to obtain the highest effectiveness with the smallest mean response time across all models. An additional efficiency gain, ranging from $1.5\times$ to $2.5\times$ relative to \pulse, is realized when utilizing the raw BM index with \pulse{}. Notably, with our proposed approach, both \esplade\ and \unicoil\ achieve sub-millisecond average response times, with a negligible loss of no more than $1\%$ in RR@10 compared to the exhaustive scenario.

\begin{table}[h]
\footnotesize
\setlength{\tabcolsep}{3pt}
\caption{The impact of varying query term pruning ratio for efficiency and effectiveness w.r.t. \splade{}. }\vspace{-3mm}
\centering
\begin{tabular}{lrrrrrrrrrr}
\toprule
\multicolumn{1}{c}{$\beta$} & $0.1$  &$0.2$&$0.3$&$0.4$&$0.5$&$0.6$&$0.7$&$0.8$&$0.9$&$1.0$ \\ 
\midrule
MRT & 0.5 & 0.9 & 1.2 & 1.5 & 1.8 & 2.2 & 2.5 & 2.8 & 3.0 & 3.2 \\
RR@10 & 30.38 & 35.81 & 37.31 & 37.78 & 38.13 & 38.12 & 38.03 & 38.04 & 38.06 & 37.97\\

\bottomrule
\end{tabular}
\label{tab:query-pruning}
\end{table}

Another pruning mechanism can be applied to query terms based on their term weights. Table~\ref{tab:query-pruning} illustrates the impact of query term pruning when executed by \pulse{} on \splade{}. \pulse{} achieves sub-millisecond retrieval times with only a slight decrease in precision.

\section{Conclusions and future work}
Learned sparse models offer significant improvements in retrieval quality while narrowing the efficiency gap between neural retrieval and quicker traditional similarity models. In this paper, we introduced \pulse, a query processing strategy designed to address the efficiency challenges inherent in learned sparse retrieval.
Future research will aim to investigate how \pulse\ compares to graph-based approximate nearest neighbor (ANN) algorithms when applied to sparse retrieval, offering potential insights into further optimizing retrieval efficiency.

\vspace{-3mm}
\subsection*{Acknowledgments}

This work is supported, in part, by the Spoke ``FutureHPC \& BigData'' of the ICSC – Centro Nazionale di Ricerca in High-Performance Computing, Big Data and Quantum Computing, the Spoke ``Human-centered AI'' of the M4C2 - Investimento 1.3, Partenariato Esteso PE00000013 - "FAIR - Future Artificial Intelligence Research", funded by European Union – NextGenerationEU, the FoReLab project (Departments of Excellence), and the NEREO PRIN project (2022AEFHAZ) funded by the Italian Ministry of Education and Research.
\balance
\bibliographystyle{acm}
\bibliography{sigir24}
\end{document}